\documentclass[10pt,letterpaper]{article}
\usepackage[top=0.85in,footskip=0.75in,marginparwidth=2in]{geometry}

\usepackage{amsmath}
\usepackage{lmodern}

\usepackage[utf8]{inputenc}

\usepackage{cite}

\usepackage{nameref,hyperref}

\usepackage[right]{lineno}

\usepackage{microtype}
\DisableLigatures[f]{encoding = *, family = * }

\raggedright
\usepackage{changepage}

\usepackage[aboveskip=1pt,labelfont=bf,labelsep=period,singlelinecheck=off]{caption}

\makeatletter
\renewcommand{\@biblabel}[1]{\quad#1.}
\makeatother

\usepackage{lastpage,fancyhdr,graphicx}
\usepackage{epstopdf}
\fancyheadoffset[L]{2.25in}
\fancyfootoffset[L]{2.25in}

\usepackage{color}

\definecolor{Gray}{gray}{.25}

\usepackage{graphicx}

\usepackage{sidecap}

\usepackage{wrapfig}
\usepackage[pscoord]{eso-pic}
\usepackage[fulladjust]{marginnote}
\reversemarginpar

\begin{document}
\vspace*{0.35in}

\begin{flushleft}
{\huge
\textbf\newline{Interactions between chronic diseases: asymmetric outcomes of co-infection at individual and population scales}}
\newline
\\
Erin E. Gorsich\textsuperscript{1,2,*},
Rampal S. Etienne\textsuperscript{3},
Jan Medlock\textsuperscript{1},
Brianna R. Beechler\textsuperscript{1},
Johannie M. Spaan\textsuperscript{2},
Robert S. Spaan\textsuperscript{4},
Vanessa O. Ezenwa\textsuperscript{5}, 
Anna E. Jolles\textsuperscript{1} \\
\bigskip
1 Department of Biomedical Sciences, 105 Dryden Hall, Oregon State University, Corvallis OR 97331 \\
2 Department of Integrative Biology, Cordley Hall, Oregon State University, Corvallis, OR, 97331 \\
3 Groningen Institute fo Evolutionary Life Sciences, University of Groningen, P.O. Box 1103, 9700 CC Groningen, The Netherlands \\
4 Department of Fisheries and Wildlife, 104 Nash Hall, Oregon State University, Corvallis, OR 97331
5 Odum School of Ecology and Department of Infectious Diseases, College of Veterinary Medicine, University of Georgia, Athens, GA, 30602
\bigskip
* eringorsich@gmail.com
\end{flushleft}

\section*{Abstract}
Co-infecting parasites and pathogens remain a leading challenge for global public health due to their consequences for individual-level infection risk and disease progression. 
However, a clear understanding of the population-level consequences of co-infection is lacking. 
Here, we constructed a model that includes three individual-level effects of co-infection: mortality, fecundity, and transmission. 
We used the model to investigate how these individual-level consequences of co-infection scale up to produce population-level infection patterns.
To parameterize this model, we conducted a four-year cohort study in African buffalo to estimate the individual-level effects of co-infection with two bacterial pathogens, bovine tuberculosis (BTB) and brucellosis, across a range of demographic and environmental contexts.
At the individual-level, our empirical results identified BTB as a risk factor for acquiring brucellosis, but we found no association between brucellosis and the risk of acquiring BTB. 
Both infections were associated with reductions in survival and neither infection was associated with reductions in fecundity. 
Results of the model reproduce co-infection patterns in the data and predict opposite impacts of co-infection at individual and population scales: whereas BTB facilitated brucellosis infection at the individual-level, our model predicts the presence of brucellosis to have a strong negative impact on BTB at the population-level.
In modeled populations where brucellosis is present, the endemic prevalence and basic reproduction number ($R_o$) of BTB were lower than in populations without brucellosis. 
Therefore, these results provide a data-driven example of competition between co-infecting pathogens that occurs when one pathogen facilitates secondary infections at the individual level.

\section*{Significance Statement}
Infection with multiple parasite species is common and the majority of co-infections involve at least one long-lasting infection. 
Our data-driven model of chronic co-infection dynamics shows that accurate prediction at the population-level requires quantifying both the individual-level transmission and mortality consequences of co-infection. 
The infections characterized in this study compete at the population-level. Competition occurs because one pathogen facilitates both the transmission and progression of the second pathogen. 
This mechanism of competition is unique compared to previously described mechanisms and occurs without of cross-immunity, resource competition within the host, or a period of convalescence.
We recommend assessing the generality of this mechanism, which could have important consequences for other chronic, immunosuppressive pathogens such as HIV or TB.


\section*{Introduction}
Over one sixth of the global human population is estimated to be affected by co-infection (concurrent infection by multiple pathogens; \cite{griffiths_nature_2011}). 
Their ubiquity includes over 270 pathogen taxa and many important chronic infections, such as hepatitis-C, HIV, TB, and schistosomiasis \cite{griffiths_nature_2011, gandhi_extensively_2006, alter_epidemiology_2006}. 
Mounting evidence suggests that co-infecting pathogens can interact within the host to influence the individual-level clinical outcomes of infection \cite{beechler_enemies_2015, graham_malaria-filaria_2005}. 
These interactions may also influence the spread of infections at the population-level \cite{abu-raddad_dual_2006, ezenwa_opposite_2015}. 
Understanding the effects of co-infection at both levels may, therefore, be fundamental to the success of integrated treatment and control programs that target multiple infections \cite{abdool_karim_integration_2011, hotez_incorporating_2006}.

One challenge to predicting the epidemiological consequences of co-infection is that the mechanisms of parasite interaction — and their resulting changes to susceptibility or disease progression — occur within the host, while patterns relevant for disease control occur within a population \cite{viney_chapter_2013}. 
Bridging these individual and population scales requires synthesizing multiple, individual-level processes across natural demographic and environmental variation. 
For example, co-infecting pathogens may be one of the best predictors of individual-level infection risk for a second pathogen \cite{lello_relative_2013, telfer_species_2010}, resulting in increased or decreased transmission. 
Co-infecting pathogens may also moderate the individual-level survival and fecundity costs of infection \cite{beechler_enemies_2015, pedersen_interaction_2008}. 
Yet, the population-level consequences of co-infection are influenced by the net effects of these potentially non-linear individual-level processes \cite{martcheva_role_2006, vasco_tracking_2007}.

At the population-level, theoretical studies have highlighted the range of dynamics generated by co-infecting pathogens \cite{abu-raddad_impact_2004, abu-raddad_dual_2006, cummings_dynamic_2005}. Even for unrelated pathogens, co-infection can dramatically modify infection dynamics through ecological mechanisms such as convalescence and disease induced mortality \cite{huang_dynamical_2005, huang_age-structured_2006, rohani_population_1998, rohani_ecological_2003, vasco_tracking_2007}. This theoretical work builds on a detailed database of childhood infections, thereby providing a data-driven understanding of co-infection dynamics for acute, immunizing infections. In contrast, data and theory on the effects of co-infection with long-lasting infections are limited  (but see, \cite{lloyd-smith_hiv-1/parasite_2008}). Chronic co-infections are of particular interest in this context, because they are responsible for the majority of co-infections \cite{griffiths_nature_2011} and have the potential to dramatically alter infection patterns \cite{martcheva_role_2006}. Their protracted presence in the host brings increased complexity to pathogen interactions, challenging model development and evaluation. Detailed longitudinal sampling or experimental studies are required to unravel their precise mechanisms and potentially asymmetric outcomes of interaction \cite{lloyd-smith_hiv-1/parasite_2008}. Few datasets simultaneously estimate the individual-level transmission, survival, and fecundity consequences of co-infection. To address this gap, we provide a data-driven investigation of co-infection dynamics for chronic pathogens.

We focused our research on two chronic bacterial infections, bovine tuberculosis (BTB) and brucellosis, in a wild population of African buffalo (/textit{Syncerus caffer}) to ask, how do the individual-level consequences of co-infection scale up to produce population-level infection patterns? This system allows us to simultaneously monitor both individual and population levels of the infection process \cite{beechler_enemies_2015, ezenwa_opposite_2015} in a natural reservoir host \cite{gomo_survey_2012, michel_mycobacterium_2010}. Furthermore, BTB and brucellosis have well-characterized and asymmetric effects on the within-host environment. BTB is a directly-transmitted, life-long respiratory infection that dramatically modifies host immunity \cite{waters_tuberculosis_2011}. African buffalo infected with BTB have reduced innate immune function and increased inflammatory responses \cite{beechler_enemies_2015}. Conversely, brucellosis is a persistent infection of the reproductive system. It persists within phagocytirc cells \cite{roop_survival_2009}, and infection invokes a less severe immune response compared to BTB \cite{Ko_molecular_2003}. These differences and our ability to observe the natural history of both infections make BTB and brucellosis an ideal system to explore disease dynamics across scales.

\begin{figure*} [hb] 
\centering
\includegraphics[width=14cm]{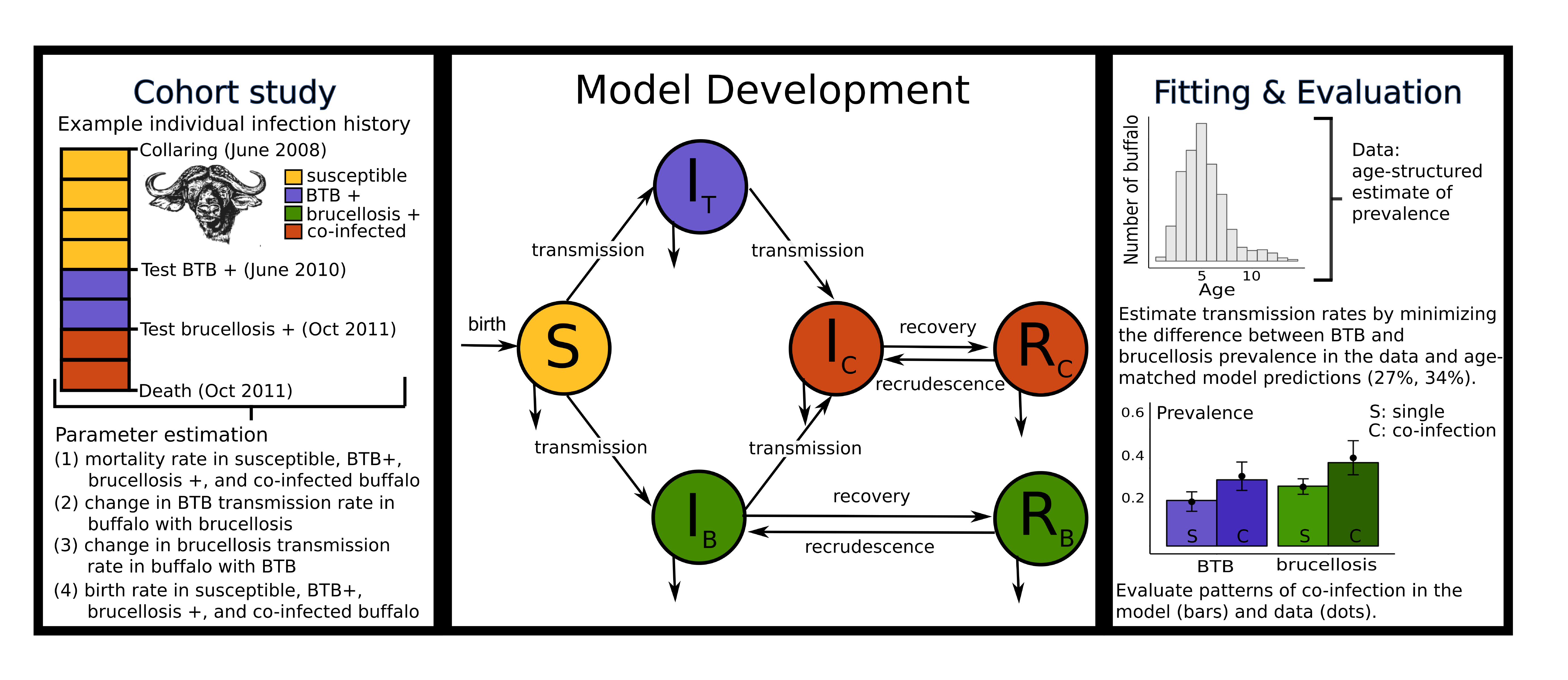}
\caption{Conceptual diagram of data collection, model, and evaluation. (center) Schematic representation of the disease model defined in SI Appendix 2. Hosts are represented as Susceptible ($S$), infected with TB only ($I_T$), infected with brucellosis only ($I_B$), co-infected with both infections ($I_C$), persistently infected with brucellosis only but no longer infectious ($R_B$), and persistently infected with brucellosis and co-infected with BTB ($R_C$). (left) A detailed cohort study informs model parameterization, including the mortality, transmission, and fecundity consequences of co-infection as well as the (right) transmission parameters for both infections. The prevalence plot illustrates that the model accurately reproduces the observed co-infection patterns in the data.  Bars represent model results and dots represent the data.}
\label{fig:fig1}
\end{figure*}

Our approach combines a novel mathematical model of the co-infection dynamics of BTB and brucellosis and a 4-year cohort study of 151 buffalo (Fig \ref{fig:fig1}). For this model, all parameters describing the consequence of co-infection were estimated from field data; they include the individual-level consequences of co-infection on mortality, fecundity, and infection risk. We quantified these parameters by tracking the individual infection profiles of each buffalo, which were monitored at approximately six-month intervals and resulted in over 4386 animal-months of observation time from two capture sites. We show that the model accurately reproduces observed co-infection patterns and use the model to predict the reciprocal effects of brucellosis and BTB on each other’s dynamics. In addition, we assess the relative importance of each individual-level process on co-infection dynamics.

\section*{Results}

\subsection*{Individual-level consequences of co-infection: model parameterization} 

\begin{figure}[ht]
\centering
\includegraphics[width=.9\linewidth]{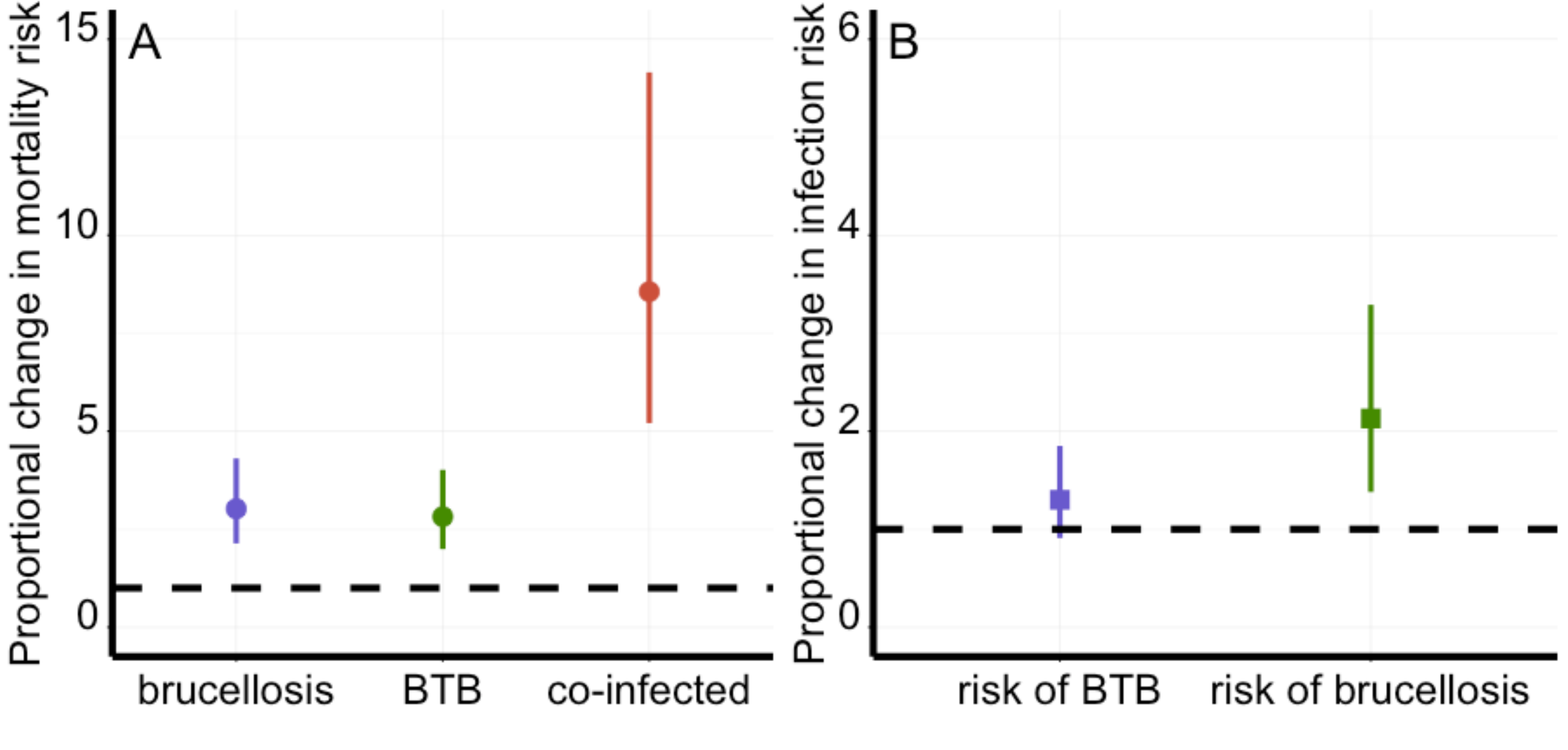}
\caption{Parameter estimation: Cox proportional hazards analysis of the cohort study. (a) Predicted mortality estimates for uninfected, infected, and co-infected adult buffalo. (b) Predicted estimates and standard error for the proportional increase in infection risk in buffalo with the second infection compared to susceptible buffalo. Brucellosis infection risk was 2.09 times higher in animals with BTB compared to susceptible animals but BTB infection risk was similar in buffalo with and without brucellosis. The dashed line indicates no change in infection risk.}
\label{fig:fig2}
\end{figure}

BTB and brucellosis were associated with additive increases in mortality (Fig \ref{fig:fig2}a; SI Appendix 1 Table S1). Approximate annual mortality rates in the data were 0.056 (10 mortalities/175.75 animal years) in uninfected buffalo, 0.108 (6 mortalities/55.5 animal years) in buffalo with BTB alone, 0.144 in buffalo with brucellosis alone (13 mortalities/ 90.5 animal years), and 0.21 (9 mortalities/43.8 animal years) in co-infected buffalo. After accounting for environmental and demographic covariates with a Cox proportional hazards regression model, BTB was associated with a 2.82 (95\% CI 1.43- 5.58) fold increase in mortality, and infection with brucellosis was associated with a 3.02 (95\% CI 1.52-6.01) fold increase in mortality compared to uninfected buffalo. Co-infected buffalo were associated with an 8.58 (95\% CI 3.20-22.71) fold increase in mortality compared to uninfected buffalo (Fig \ref{fig:fig2}a). Mortality rates were also influenced by buffalo age and capture site, but the effect of co-infection remained consistent across all ages and in both sites (SI Appendix 1). Neither infection was associated with reductions in fecundity (described in detail in SI Appendix 1, Fig S1). Uninfected buffalo were observed with a calf 68\% (11/16) of the time compared to 37\% (6/16), 29\% (7/24), and 57\% (4/7) in BTB positive, brucellosis positive, and co-infected adult buffalo.

The consequences of co-infection on infection risk were asymmetric (Fig \ref{fig:fig2}b; SI Appendix 1, Table S2). Approximate brucellosis incidence rates were 0.05 (18 infections/340 animal years) in uninfected buffalo compared to 0.08 (8 infections/104 animal years) in buffalo with BTB. Approximate BTB incidence rates were 0.08 (27 infection/ 340 animal years) in uninfected buffalo and 0.07 (9 infections/ 138 animal years) in buffalo with brucellosis. After accounting for demographic covariates in a Cox proportional hazards regression model, brucellosis infection risk was 2.09 (95\% CI 0.89 – 4.91) times higher in buffalo with BTB compared to susceptible buffalo. BTB infection risk was similar in uninfected buffalo and buffalo with brucellosis. The effect BTB on brucellosis infection risk varied by capture-site, with the effect of BTB varying from no change at one site to a 4.32 (95\% CI 1.51 – 12.37) fold increase in risk at the other site (interaction term for BTB$\times$site: p-value = 0.045; SI Appendix 1, Table S2). Brucellosis infection risk also varied with buffalo age. Early reproductive-aged buffalo were associated with increased infection risk. Our model parameterization, therefore, represents increased brucellosis transmission in early reproductive-aged buffalo and the average effect of BTB on brucellosis infection risk accross sites. Talbes S3 and S4 in SI Appendix 2 provides a summary of the consequences of co-infection on brucellosis transmission, BTB transmission, mortality, and fecundity quantified in our data analyses.

\subsection*{Population-level consequences of co-infection: basic reproduction number and prevalence}

\begin{figure}[hb]
\centering
\includegraphics[width=.8\linewidth]{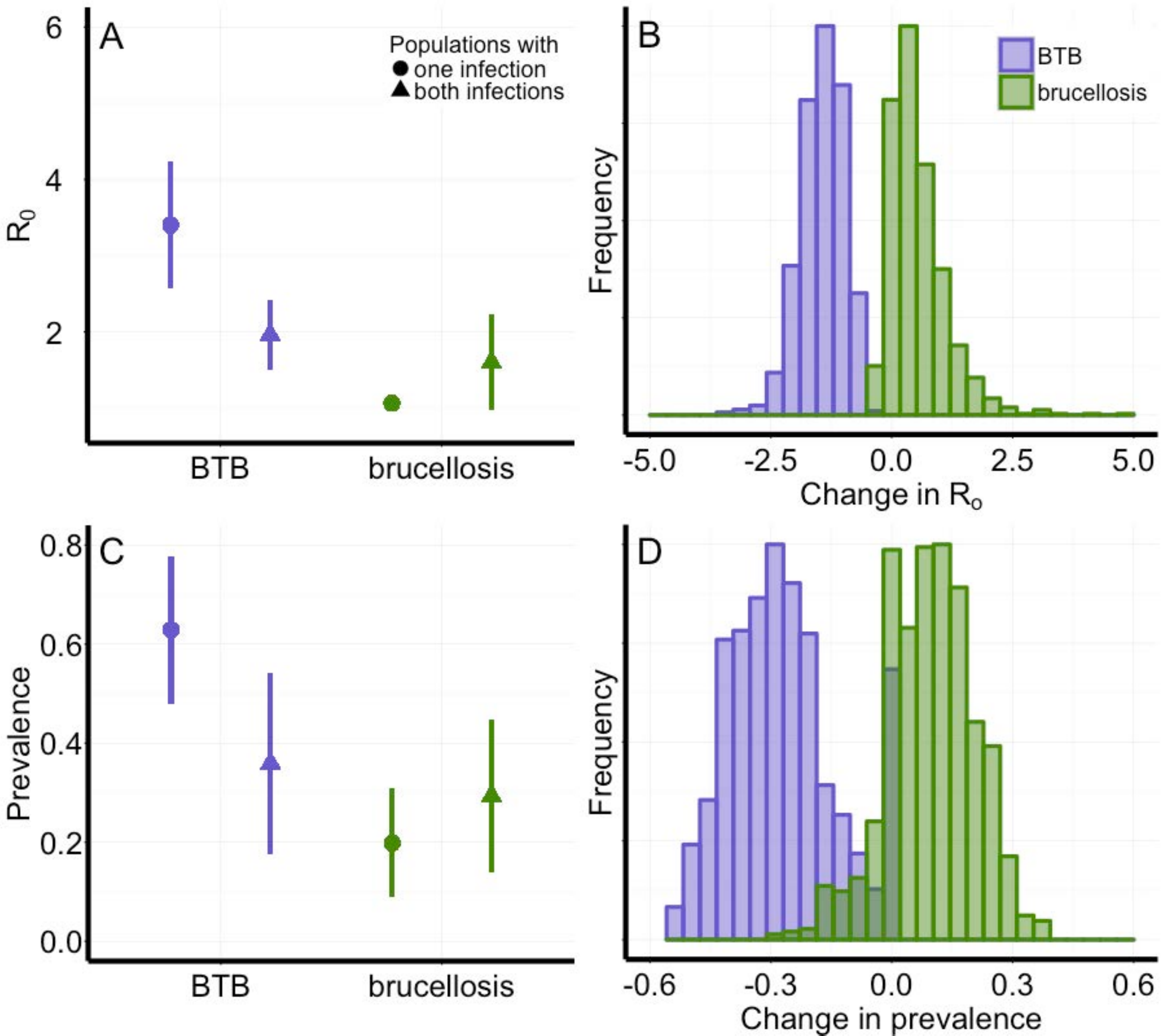}
\caption{Model predictions: reciprocal consequences of co-infection for the basic reproduction number ($R_o$) and endemic prevalence. (a) The estimated $R_o$ for BTB was lower in populations where brucellosis co-occurs while the estimated $R_o$ for brucellosis was similar in populations with and without BTB. (b) The estimated endemic prevalence of BTB was lower in populations where brucellosis co-occurs while the estimated endemic prevalence of brucellosis was similar in populations with and without BTB.  Uncertainty was incorporated into model predictions using Monte Carlo sampling of parameters describing the individual-level consequences of co-infection (SI Appendix 3). Dots and lines represent the mean and standard error of model predictions. (c) Histograms of the difference in $R_o$ and (d) endemic prevalence show the negative consequence of co-infection for BTB and neutral consequences of co-infection for brucellosis at the population-level. Change is calculated for each parameter combination as the predicted value in populations with co-infection subtracted by the predicted value in populations with a single pathogen. Purple bars and lines represent BTB; green lines represent brucellosis. }
\label{fig:fig3}
\end{figure}

We built a disease dynamic model to translate the individual-level effects quantified above into predicted population-level effects of co-infection (Fig \ref{fig:fig1}; SI Appendix 2). We estimated parameter values for the transmission rate of BTB and brucellosis by minimizing the sum of squared errors between BTB and brucellosis infection prevalence in our data and in an age-matched sample from the model. BTB prevalence in the data was 27\% and brucellosis prevalence was 34\%. The resulting transmission parameters (BTB $1.331 \times 10^{-3}$, brucellosis $5.764 \times 10^{-1}$) accurately predict the positive association between BTB and brucellosis observed in the data (Fig \ref{fig:fig1}) and allow us to predict equilibrium infection levels and the basic reproduction number in modeled populations with and without co-infection. The basic reproduction number, $R_{o,i}$ (i = T, B for infection with BTB or brucellosis) is defined as the average number of secondary cases generated by a single infection in a susceptible population. The presence of brucellosis infection results in large reductions in $R_{o,T}$, with a predicted $R_{o,T}$ = 3.4 in populations where brucellosis is absent and $R_{o,T}$= 1.5 in populations where brucellosis is present. The predicted equilibrium BTB prevalence was also lower in populations where both pathogens occur, with a BTB prevalence of 65.8\% in populations where brucellosis is absent compared to 27.9\% when both pathogens co-occur. Conversely, the presence or absence of BTB has only minor effects on the $R_o$ and equilibrium prevalence of brucellosis. To represent uncertainty in the individual-level consequences of co-infection, we used Monte Carlo sampling of the parameters quantified in our statistical analyses (Fig \ref{fig:fig2}; SI Appendix 2).  Figure \ref{fig:fig3} displays the effect of co-infection when uncertainty in input parameters is considered. In this range of parameter values (parameter space), 96\% of model trajectories predicted a lower BTB prevalence in populations with co-infection. In the remaining 4\%, brucellosis did not persist in populations with or without co-infection due to high mortality rates and low facilitation rates (Fig S4, SI Appendix 2).

\begin{figure}
\centering
\includegraphics[width=.99\linewidth]{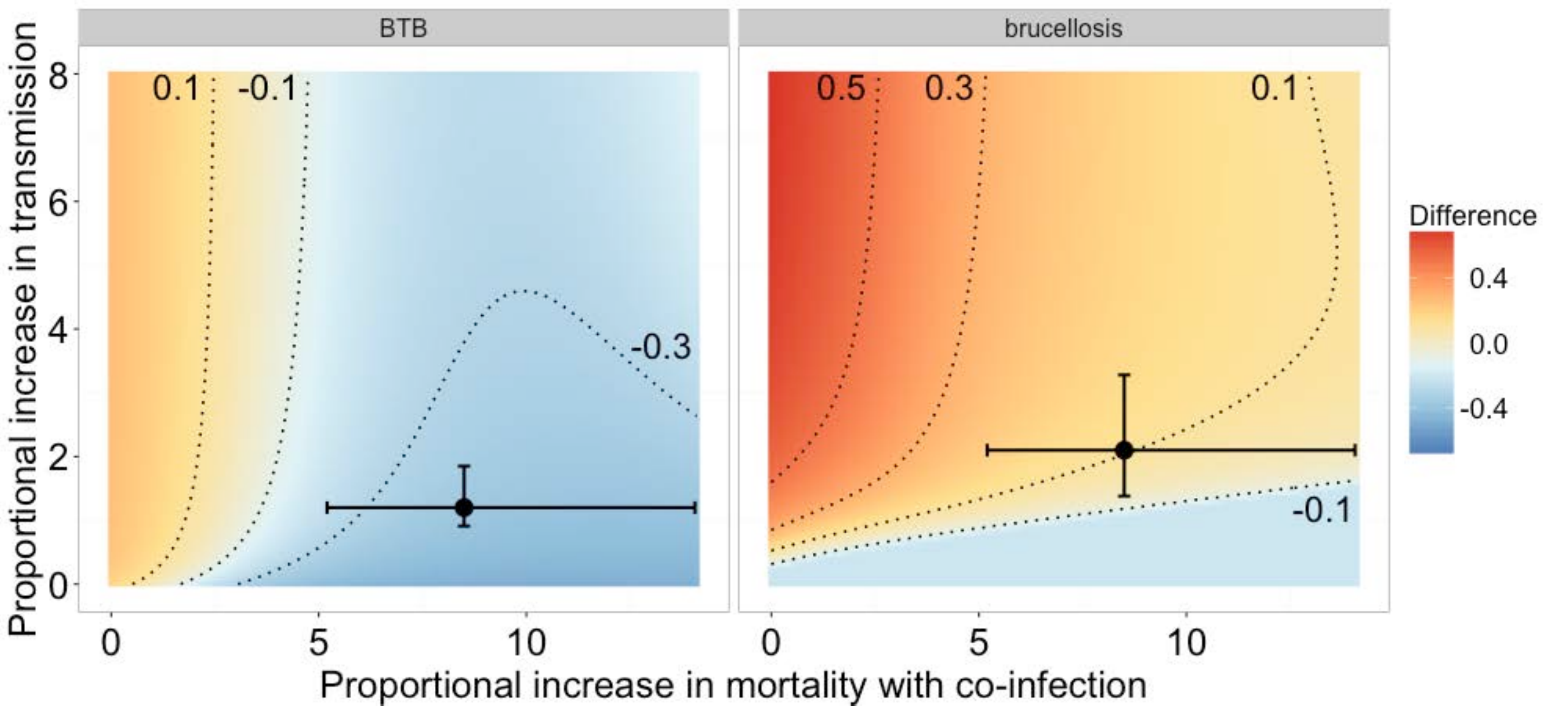}
\caption{The difference between predicted prevalence values in populations with and without co-infection for BTB prevalence (left panel) and brucellosis prevalence (right panel). Reds indicate that infection with the second pathogen increased in the presence of the focal pathogen; blues indicate that infection with the second pathogen decreased in the presence of the focal pathogen; yellows indicate no change. Contour lines indicate changes in prevalence by 20\%. Stars and error bars indicate mean and standard error parameter values estimated in the data. Proportional increases in mortality are measured as the increase compared to the rate in susceptible individuals. Proportional increases in transmission represent the increase in the transmission rate for the focal pathogen in the presence of the co-infecting pathogen.}
\label{fig:fig4}
\end{figure}

To generalize these results, we explored the population-level consequence of co-infection over a range of parameter values (\ref{fig:fig4}). By manipulating the transmission and mortality consequences of co-infection, we explored infection levels in other environmental contexts where the individual effects of co-infection may be reduced or exacerbated. The resulting prevalence surfaces suggest that, co-infecting pathogens will have a negative effect on prevalence if co-infected individuals have elevated mortality and reduced or similar susceptibility. Co-infecting pathogens are predicted to have a positive effect if co-infected individuals have increased susceptibility and minimal changes in mortality. In contrast, BTB and brucellosis are predicted to have drastically different responses to co-infection for parameter values where co-infection modifies both processes. BTB prevalence was lower in populations with brucellosis for most parameter values while the effect of BTB on brucellosis was more variable. This difference is due to BTB’s long infection duration because the cumulative effect of facilitation increases with infection duration.

At the parameter values quantified in our empirical dataset, these results illustrate that the lower BTB prevalence in populations where brucellosis co-occurs is driven two mechanisms: (1) BTB increases the transmission rate of brucellosis but not vice versa and (2) co-infection results in increased mortality. As a result, at the individual-level, buffalo infected with BTB are more likely to become infected with brucellosis and die than their uninfected counterparts. The resulting reductions in infection duration mean that the prevalence of brucellosis is predicted to reduce BTB infection-levels at the population-level. These results are robust to several important changes in the model structure, including alternative forms of density dependence and a range of model parameters (see sensitivity analyses in SI Appendix 2).  Model dynamics in all formulations are qualitatively similar, although there is some variation in overall magnitude of change with co-infection.

\section*{Discussion}
Our study provides a mechanistic understanding of how chronic co-infections mediate each other’s dynamics. Model dynamics show that co-infection can increase or decrease the prevalence of a second pathogen depending on the net effect of co-infection on transmission and infection duration, via mortality. When co-infection modifies only transmission or only mortality, the prevalence of the second pathogen predictively increases or decreases. Previous work has quantified the disease dynamic consequences of changes in transmission through a range of mechanisms: cross immunity, antibody-mediated enhancement, immunosuppression, and convalescence \cite{abu-raddad_impact_2004, rohani_population_1998, ferguson_effect_1999, bhattacharyya_cross-immunity_2015}. Here, we show that transmission and mortality should be considered concurrently, following theoretical predictions \cite{martcheva_role_2006, lloyd-smith_hiv-1/parasite_2008, fenton_dances_2013}. When a co-infecting pathogen modifies both processes, non-linear responses mediated through the co-infecting pathogen can have a large impact on population-level disease dynamics.

By exploring the co-infection dynamics of BTB and brucellosis, we also provide a data-driven example of competition between pathogens in a natural population. Here, the mechanism driving competition is different from previously described examples that focus on cross-immunity \cite{bhattacharyya_cross-immunity_2015}, resource competition within the host \cite{graham_ecological_2008}, or ecological competition by convalescence \cite{rohani_population_1998, rohani_ecological_2003}. Specifically, the mechanism of parasite interaction in these examples occurs when one infection reduces the transmission of the second pathogen, typically by removing individuals from the susceptible pool during co-infection. By contrast, BTB facilitates the transmission of brucellosis at the individual-scale (Fig \ref{fig:fig2}). Because co-infection is associated with elevated mortality, co-infected individuals are also removed from the population at a faster rate. Competition, therefore, occurs at the population-level: BTB is predicted to have a lower prevalence and lower $R_{o,T}$ in populations where brucellosis occurs compared to populations without brucellosis.

The model structure presented in this study was informed by our empirical data. As a result, it incorporates realistic age-specific transmission and mortality rates as well as data-driven estimates of the consequences of co-infection. However, additional detail could be added to our model. Specifically, we do not know the consequences of co-infection on recovery or infectiousness, two processes likely to influence persistent infections \cite{huang_dynamical_2005, lloyd-smith_hiv-1/parasite_2008}. We also do not consider genetic variation within our buffalo population that may mediate susceptibility to both pathogens. However, our model’s ability to re-create co-infection patterns with the mechanisms characterized is encouraging. Furthermore, our empirical results account for natural variation in demographic and environmental conditions and model-predictions including this variation remain informative. Thus, our results suggest the importance of co-infection in generating population-level association patterns relative to environmental or genetic drivers of infection.

Given the ubiquity and documented individual-level impacts of chronic co-infections on the host, these results illuminate two core challenges in the design and application of integrated control strategies. First, it remains unclear how commonly competition between co-infecting pathogens is occurring. Understanding which pathogens may be competing in co-infected host populations is crucial to estimating the costs and benefits of disease control interventions. For example, in the presence of pathogen competition, removing one pathogen may unintentionally lead to a resurgence of or increases in prevalence of a competing pathogen. Our results suggest that competition at the population-level can occur between unrelated pathogens and in the absence of competition for shared resources within the host. Competition appears to be strongest when pathogens have asymmetric effects on transmission. Similar asymmetries in transmission occur in HIV-malaria \cite{abu-raddad_dual_2006} and HIV-HCV co-infections \cite{urbanus_hepatitis_2009}, suggesting a role for this mechanism in other systems. Second, knowledge on which chronic pathogens are most likely to be influenced by co-infection remains largely theoretical (excluding notable progress with HIV- co-infection \cite{abu-raddad_dual_2006, lloyd-smith_hiv-1/parasite_2008}). Here, the immunosuppressive pathogen, BTB \cite{beechler_innate_2012, beechler_enemies_2015}, was strongly influenced by co-infection at the population-level, and our analyses show that BTB prevalence should typically decline in the presence of another chronic pathogen, provided that co-infected hosts suffer greater mortality. This raises the question of whether there are traits of chronic pathogens (e.g. immunosuppressive effects) that make them more likely to be influenced by co-infection. Studies addressing these questions are urgently needed to target both research and treatment on the pathogens most likely to be influenced by co-infection. Such a research agenda will require additional synthesis of theoretical and empirical work: theoretical studies are needed to characterize how population-level infection dynamics will response to changes in transmission and mortality for different types of pathogens and empirical studies are needed to quantify when and how these parameters change through pathogen interactions. In combination, this work can provide clarity about the consequences of co-infections, and their control, at both the individual and population scales.

\section*{Materials and Methods}
\subsection*{Model Development}
We developed an age-structured continuous time disease dynamic model to explore the consequences of co-infection on BTB and brucellosis infection (Fig \ref{fig:fig1}).
Animals are represented with six groups: susceptible to both infections (S), infected with BTB only (IT), infected with brucellosis and infectious (IB), co-infected with both pathogens (IC), persistently infected with brucellosis but no longer infectious (RB), or persistently infected with brucellosis and co-infected with BTB (RC). We modeled BTB as a lifelong infection with density dependent transmission \cite{jolles_interactions_2008}. 
Transmission of brucellosis was assumed to be frequency dependent because transmission occurs through ingestion of the bacteria shed in association with aborted fetuses, reproductive tissues, or discharges during birthing \cite{samartino_pathogenesis_1993}. 
We did not consider vertical transmission because serological evidence suggests that it is rare in African buffalo \cite{gorsich_context-dependent_2015} and experimental evidence for vertical transmission varies by host species (cattle \cite{fensterbank_congenital_1978}, bison \cite{plommet_brucellose_1973}, elk \cite{thorne_brucellosis_1978}). Buffalo populations experience density dependent recruitment \cite{sinclair_resource_1975}, represented with a generalized Beverton and Holt equation \cite{getz_hypothesis_1996}.
This 2 parameter representation of density dependence gives a stable age structure and relatively constant population size (\cite{cross_assessing_2006}; SI Appendix 2, Fig S3). A full description of the model is provided in SI Appendix 2.

The individual-level consequences of co-infection can be summarized by 4 individual-level processes: (1) the effects of prior infection with brucellosis on the rate of acquiring BTB infection (2) the effects of prior infection with BTB on the rate of acquiring brucellosis infection, (3) the effects of co-infection on mortality rate, and (4) the effects of co-infection on birth rates. 
To investigate the consequences of these four individual level processes on disease dynamics, we quantified the mean values of these rates in susceptible, singly infected, and co-infected buffalo. 
Transmission rates, mortality rates, and the proportional reductions in fecundity with infection may be age dependent, but recovery and recrudescence are assumed to be independent of age.

\subsection*{Individual-level data and parameter estimation}
We conducted a longitudinal study of 151 female buffalo to estimate the consequences of BTB and brucellosis infection. Buffalo were captured at two locations in the south-eastern section of KNP, radio collared for re-identification, and re-captured biannually at approximately 6 month intervals until June-October of 2012.  During each capture, we recorded brucellosis infection status, BTB infection status, age, and the animals’ reproductive status. Brucellosis testing was conducted with an ELISA antibody test and BTB testing was conducted with a gamma-interferon assay  \cite{gorsich_evaluation_2015, michel_approaches_2011}. Detailed methodological descriptions of our capture and disease testing protocols are provided in SI Appendix 3.

We assessed the effects of co-infection on mean mortality rates and the mean rate animals acquired infection by analyzing our longitudinal time-to-even data using semi-parametric Cox models where an individual’s covariates representing infection change over time. Specifically, we fit three regression models to predict three events: the time-to-mortality in uninfected, BTB+, Brucellosis +, and co-infected individuals; the time-to-infection with brucellosis in buffalo with and without BTB; and the time-to-infection with BTB in buffalo with and without brucellosis. In all analyses, we include age and initial capture site as time-independent, categorical variables and infection status as a time-dependent explanatory variable. We also evaluate whether the association between brucellosis and BTB varied by age or site by including interactions terms between BTB and each environmental variable. The supplemental information includes a detailed quantification of our re-capture rates, justifies the use of the Cox model, and outlines our model selection procedure. 

\subsection*{Model evaluation and inference}
Parameter values for the transmission rate of BTB and brucellosis were estimated by fitting the model to prevalence estimates in the study population. Our data do not represent a random sample because buffalo aged over the course of the study, with a median age of 3.4 years in buffalo initially captured in June-October 2008. We, therefore,  calculate prevalence after randomly sampling one time point for each individual.  We estimate prevalence in the study population as the mean prevalence in 1000 replicate samples. Model estimates of prevalence were calculated numerically using the deSolve package \cite{desolve_package}. The transmission rates of both pathogens were estimated by numerically minimizing the sum-of-squared errors between the mean prevalence estimates for BTB and brucellosis and an age-matched estimate of prevalence from the model. We used the Nelder-Mead algorithm implemented with the optim function in R to minimized this function. We evaluated our model by comparing its ability to recreate co-infection patterns in the data (Fig \ref{fig:fig1}). We calculated Ro numerically using the next generation method \cite{van_den_driessche_reproduction_2002}, reviewed in \cite{heffernan_perspectives_2005}. We calculated endemic infection levels of both pathogens in populations with and without the co-infecting pathogen.

\section*{Supporting Information}
Extended methods are provided in the SI Appendix. \\
\noindent{}Appendix 1: Additional information statistical analysis \\
\noindent{}Appendix 2: Additional information on model development and analysis \\
\noindent{}Appendix 3: Additional information on field methods and diagnostic testing \\


\section*{Acknowledgments}
We thank South African National Parks (SANParks) for their permission to conduct this study in Kruger. We thank P. Buss, M. Hofmeyr and the entire SANPark’s Veterinary Wildlife Services Department for animal capture and logistical support. We thank the Webb lab group for comments on the manuscript and technical support. Animal protocols for this study were approved by the University of Georgia (UGA) and Oregon State University (OSU) Institutional Animal Care and Use Committees (UGA AUP A2010 10-190-Y3-A5; OSU AUP 3822 and 4325). This study was supported by a National Science Foundation Ecology of Infectious Diseases Grant to A. Jolles and V. Ezenwa (EF-0723918/DEB-1102493, EF-0723928) and a NSF-GRFP and NSF-DDIG award to E. Gorsich (DEB-121094).
\nolinenumbers

\pagebreak

\begin{center}
\textbf{\Large Supporting Information: \\
Interactions between chronic diseases: asymmetric outcomes of co-infection at individual and population scales}
\end{center}

\noindent \Large{\textbf{Appendix 1. Additional information on statistical analyses}}\\
\normalsize

{\subsection*{Mortality}}
Our mortality dataset included 127 individuals and 38 mortality events. We estimated the time of mortality as midway between the last capture period the animal was observed and the subsequent capture period (6 months later) when it was recovered unless the exact mortality date was known. This assumption is supported by
our sampling design that recovered most animals quickly after death. For 20 of the mortality events included here, we were able to estimate the mortality date within one month of error by comparing the date the animal was last observed with the date it was known to be dead. Animals were identified as dead by either finding the carcass or recovering the radio-collar. All radio-collars were recovered before the subsequent capture period. The high recapture rates in this study made the Cox proportional hazards model an acceptable representation of the time-to-event process \cite{fisher_time-dependent_1999}. We evaluated the proportional hazards assumption of the model using Schoenfeld residuals \cite{grambsch_proportional_1994, fox_cox_2011} quantified with the cox.zph function in the survival package \cite{therneau_package_2014}.
To test for differences in the distribution of mortality times in uninfected and infected buffalo, we included the presence or absence of brucellosis-positive test result ($bruc_i$) and a BTB-positive test result ($BTB_i$) as time-dependent explanatory variables. We also account for the buffalo's age at first capture ($agecat_i$) and capture location ($site_i$) as time-independent, categorical variables. These main effects and interaction terms give the following full model:
\begin{gather*}
h_{i}(t) \sim h_{0}(t) \text{exp} (\beta_{1} BTB_i + \beta_{2} bruc_i + \beta_{3,j}agecat_i + \beta_4 site_i + \\
\beta_5 BTB_i * bruc_i + \beta_{6,j} BTB_i * agecat_i + \beta_{7,j} bruc_i * agecat_i + \\
 \beta_8 BTB_i * site_i + \beta_9 bruc_i * site_i)
\end{gather*}

where $h_{i}(t)$ represents the hazard function for the i th individual and describes the instantaneous risk of mortality at time, t, conditional on survival to that time. The baseline hazard is represented by $h_{o}(t)$ and remains unspecified.  

We conducted model selection in two stages. First, we considered the relationship between the buffalo's age at first capture and mortality rate by only considering age as a predictor of mortality. This is because both infections are assumed to be life-long, so we were concerned that the positive association between age and testing positive for both infections could bias our results. We considered 4 potential representations of age as an explanatory variable (age1: late development with a sub-adult category, 1-2.9, 3-5.9, 6+; age2: late development without a sub-adult category, 1-2.9, 3+; age3: early development with a sub-adult category, 1-1.9, 2-4.9, 5+; age4: early development without a sub-adult category, 1-1.9, 2+). The age categories representing late development (age1: 1-2.9, 3-5.9, 6+, age2: 1-2.9, 3+) provided better fits based on AIC (Akaike information criterion) compared to the age categories representing early development with the same number of parameters (age3: 1-1.9, 2-4.9, 5+, age4: 1-1.9, 2+). The addition of an extra age category had a minimal effect on model fit. The AIC values for each model were, 330.9, 329.9, 336.6, and 335.8, for models including age1, age2, age3, and age4, respectively. Second, we conducted a formal model selection following the full model in equation 1 with age2 representing the covariate for age. We used backwards selection by sequentially removing interaction terms and then main effects terms based on drop-in-deviance $\chi^2$ tests. From the final model, we evaluate $\beta_1$ to determine the mortality consequences of BTB, $\beta_2$ to determine the mortality consequences of brucellosis, and $\beta_5$ to determine if buffalo co-infected with both pathogens have exacerbated mortality.

We converted these parameters into infection-specific mortality rates by interpreting the exponentiated model coefficients as having a multiplicative effect on the mortality hazard (e.g. the mortality rate in buffalo with BTB is equal to the mortality rate in susceptible buffalo multiplied by exp($\beta_1$)).  \\

\begin{table}[hb]
\centering
\caption*{\textbf{Table S1.} Parameter values ($\beta$), standard errors (SE), and significance tests (Z-value, p-values) for each covariate in the final model. A positive parameter value for BTB or brucellosis indicates an increased mortality hazard by a factor of exp($\beta$). A positive parameter for site indicates higher mortality in the Crocodile Bridge capture site compared to Lower Sabie. A positive parameter for age indicates higher mortality in juveniles (age $<3$) compared to adults (age $\geq 3$).  The data contained 127 animals and 38 mortality events.}
\newcommand{\head}[1]{\textnormal{\textbf{#1}}}
\normalsize
\begin{tabular}{lcccc} 
\hline
\head{Parameter     } & \head{     Estimate ($\beta$)     } & \head{     SE     } & \head{     Z-value     } & \head{     p-value     } \\*
\hline
BTB ($\beta_1$) & 1.04 & 0.35 & 2.98 & 0.003 \\
bruc ($\beta_2$)  & 1.11 & 0.35 & 3.16 & 0.002\\
age ($\beta_3$) & 1.18 & 0.33 & 3.35 & < 0.001\\
site ($\beta_4$) & 0.74 & 0.32 & 2.27 & 0.023\\
\hline 
\end{tabular}
\end{table} 
 
\pagebreak 

\subsection*{Fecundity}
We assessed the effects of infection and co-infection on fecundity by testing for an association between infection status and reproductive status, defined as having a calf. Birthing in African buffalo occurs seasonally, with most births occurring from December to March (5). Buffalo calves are often weaned approximately 5-6 months later \cite{winthrop_bovine_1988}. 
We, therefore, expected calf observations to be less accurate just before or during birthing season and account for this by only including captures immediately following the birthing season (March to July). Calving is strongly age-dependent and only buffalo 4 years of age or older were considered in this analysis, following previous analyses of buffalo fecundity \cite{gorsich_context-dependent_2015}. 
The resulting dataset includes 143 observations from 62 buffalo over the course of our longitudinal study. 
Of the 143 observations, 29\% were observed with a calf at heel. 
Uninfected adult buffalo (age $\geq$ 5) were observed with a calf at heel 68\% (11/16) of the time, buffalo with BTB were observed with a calf at heel 37\% (6/16) of the time, buffalo with brucellosis were observed with a calf at heel 29\% (7/24) of the time, and co-infected adults were observed with a calf at heel 57\% (4/7) of the time.

We tested if reproductive status is associated with infection status with a generalized linear mixed model with binomial errors and a logit link function for the response.  
We considered all main effects and interactions specified for previous analyses as dependent variables: BTB infection, brucellosis infection, age, capture site and interactions between BTB-brucellosis, BTB-age, BTB-site, brucellosis-age, and brucellosis-site. 
Animals were sampled over time, so we accounted for animal ID as a random effect.

We conducted model selection on the fixed effects in two stages.
First, we considered the relationship between the buffalo's age and fecundity by only considering age as a predictor. 
In this analysis, the age categories representing early development (age3 = 1-1.9, 2-4.9, 5+, AIC = 166.5) provided better fits compared to age categories representing late development (age1 = 1-2.9, 3-5.9, 6+, AIC = 168.5). 
Second, we conducted a formal model selection following the full model with age3 representing the covariate for age. 
Model selection was conducted by sequentially removing interaction terms and then main effects terms based on drop-in-deviance $\chi^2$ tests.
After model selection, only age was a significant predictor of fecundity (Z = 3.30, p-value = 0.001).  The odds of a buffalo being observed with a calf were 3.84- fold (95\% CI 1.73- 8.54) higher in adult buffalo aged over 5 years old compared to sub-adult buffalo aged less than 5 years.
 
 \pagebreak

\begin{figure}[ht]
\centering
\includegraphics[width=.99\linewidth]{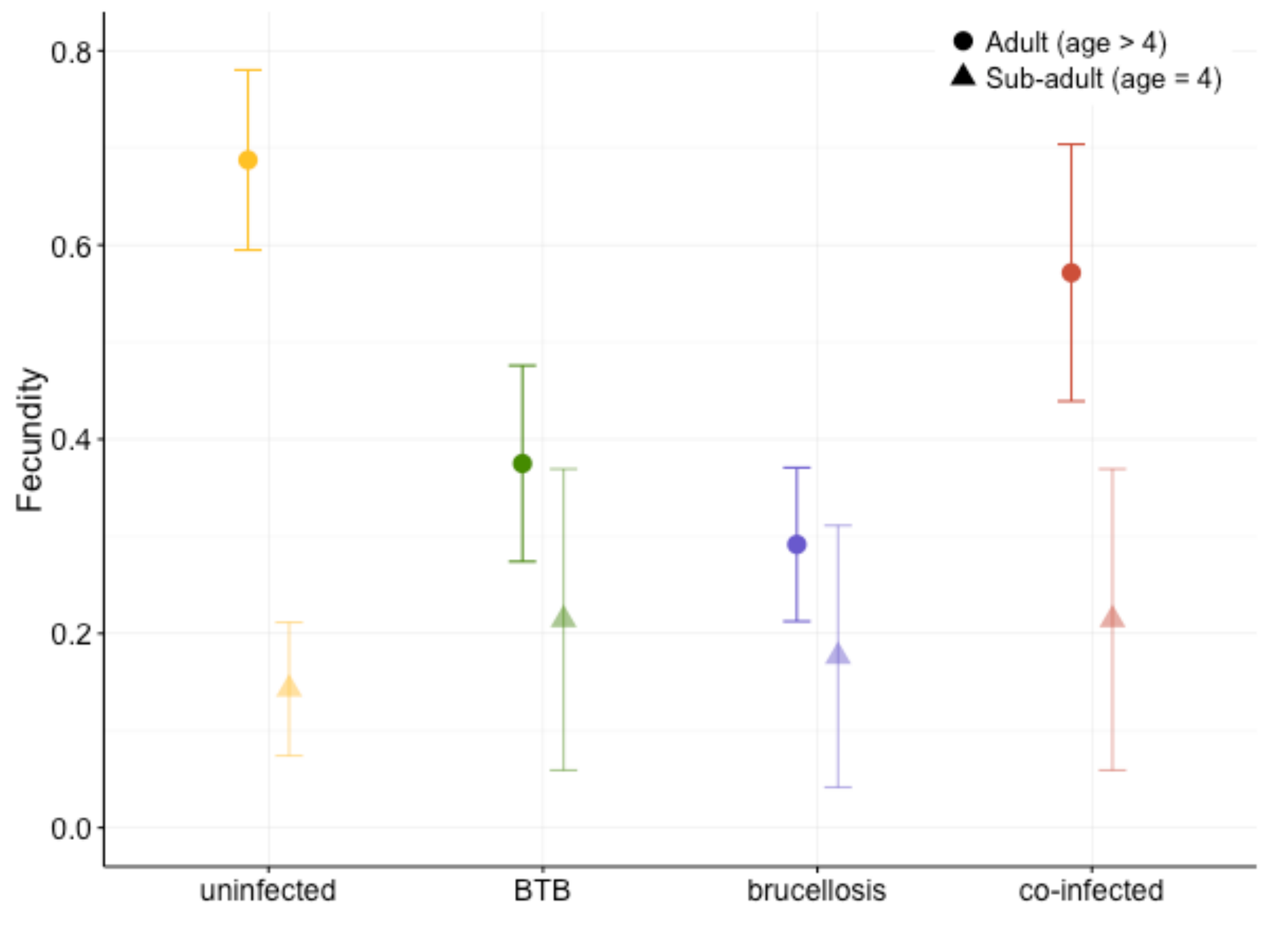}
\caption*{\textbf{Fig S1.} Proportion of buffalo observed with a calf.  Proportion estimates and standard errors are based on the number of observations with a calf in the data.}
\label{fig:figS1}
\end{figure}

\pagebreak

\subsection*{Infection risk}
Our dataset for brucellosis incidence included 26 brucellosis incidence events in the 110 buffalo who were sero-negative at their first capture (Fig S2). Our dataset for BTB incidence included 36 BTB incidence events in 131 buffalo. We used Cox proportional hazards models to test for differences in the distribution of infection times in uninfected buffalo and buffalo that tested positive for the second infection. Using the same co-variates and model selection procedure described in Table S1, we fit the following full model,

\begin{gather*}
h_{i}(t) \sim h_{0}(t) \text{exp} (\beta_{1} infection_i + \beta_{2,j} agecat_i + \beta_3 site_i \\
+ \beta_{4,j} infection_i agecat_i + \beta_{5} infection_i site_i 
\end{gather*}

where $h_{i}(t)$ represents the hazard function for the $i^{th}$ individual and describes the instantaneous risk of infection at time, t, conditional on the individual not becoming infected until that time. For models predicting brucellosis infection risk, we evaluate $\beta_1$ to estimate the consequences of BTB infection on brucellosis infection risk, $\beta_4$ to determine if the consequences of BTB vary across age groups, and $\beta_5$ to determine if the consequences of BTB vary between sites. Similarly, for models predicting BTB infection risk, we evaluate $\beta_1$ to estimate the consequences of brucellosis infection on BTB infection risk, $\beta_4$ to determine if the consequences of brucellosis vary across age groups, and $\beta_5$ to determine if the consequences of brucellosis vary between sites.

For both infections, the age categories representing late development (age1 = 1-2.9, 3-5.9, 6+, age2 = 1-2.9, 3+) provided better fits compared to the age categories representing early development with the same number of parameters (age3 = 1-1.9, 2-4.9, 5+, age4 = 1-1.9, 2+). However, we were unable to consider age1 as a co-variate because none of the 8 buffalo aged 6 years or older at the initial capture became infected with brucellosis.  AIC values for were 248.03, 250.04, and 251.63 for models predicting brucellosis infection with age2, age3, and age4 included as a predictor, respectively. AIC values were 347.80, 349.37, and 349.26 for models predicting BTB infection with age2, age3 and age4 included as a predictor. Thus, we conducted a formal model selection for brucellosis and BTB incidence following the full model with age2 representing the covariate for age. 

 We note that the effect of BTB infection on brucellosis incidence varied by capture site ($\beta_5$, BTB:site). Buffalo infected with BTB were associated with a 4.32 (95\%CI 1.51-12.37) times higher rate of brucellosis infection at the Lower Sabie site while buffalo infected with BTB at the Crocodile Bridge site had similar incidence rates compared uninfected buffalo (Z = - 0.821, p-value = 0.412). The parameters for our disease model are based on the effect of BTB taken across both sites: buffalo infected with BTB faced a 2.09 (95\% CI 0.89-4.91) fold higher risk of acquiring brucellosis (Z = 1.742, p-value = 0.081).

\begin{table}[ht]
\centering
\caption*{\textbf{Table S2.} Parameter values, standard errors, and significance tests (Z-value, p-values) are shown for each covariate in the final model. A positive parameter value for BTB or brucellosis indicates an increased mortality hazard by a factor of exp$(\beta)$. A positive parameter for site indicates higher infection risk in Crocodile Bridge capture site compared to Lower Sabie. A positive parameter for age indicates a higher infection risk in juveniles (age $<3$) compared to adults (age $\geq$ 3). }
\newcommand{\head}[1]{\textnormal{\textbf{#1}}}
\normalsize
\begin{tabular}{lcccc} 
\hline
\head{Parameter     } & \head{     Estimate ($\beta$)     } & \head{     SE     } & \head{     Z-value     } & \head{     p-value     } \\*
\hline
\textbf{Brucellosis incidence} & & & & \\
BTB ($\beta_1$) & 1.46 & 0.54 & 2.73 & 0.006 \\
age ($\beta_2$) & 1.89 & 0.43 & 2.06 & 0.039\\
site ($\beta_3$) & 0.71 & 0.51 & 1.41 & 0.158 \\
BTB:site ($\beta_5$) & -2.40 & 1.20 & -2.01 & 0.045 \\
 & & & & \\
\textbf{BTB incidence}  & & & & \\
age ($\beta_2$) & -0.61 & 0.37 & -1.65 & 0.10 \\
site ($\beta_3$) & -0.70 & 0.35 & -2.02 & 0.04\\
\hline 
\end{tabular}
\end{table}

\begin{figure}
\centering
\includegraphics[width=.99\linewidth]{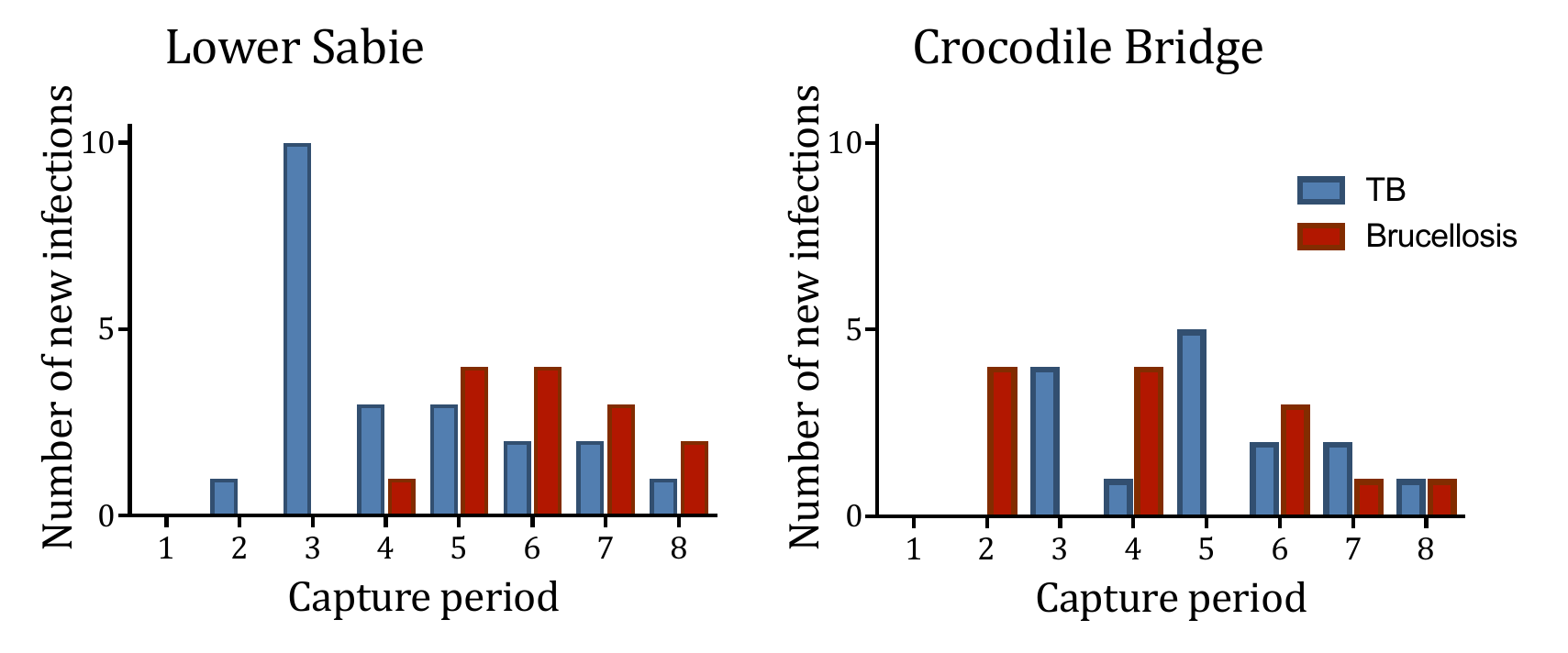}
\caption*{\textbf{Fig S2.} The number of new BTB and brucellosis infections over the course of the study in buffalo initially captured in the Lower Sabie and Crocodile Bridge sites show that infections occurred continuously over the study period.}
\label{fig:figS2}
\end{figure}

\pagebreak
\clearpage

\noindent \Large \textbf{Appendix 2. Additional information on model development and analysis}\\

\normalsize

We developed an age-structured continuous time disease dynamic model of BTB and brucellosis co-infection. 
Animals are represented with six groups: susceptible to both infections, $S(t, a)$; infected with BTB only, $I_T (t, a)$; infected with brucellosis and infectious, $I_B (t, a)$; co-infected with both pathogens, $I_C (t, a)$; persistently infected with brucellosis but no longer infectious, $R_B (t,a)$; or persistently infected with brucellosis and co-infected with BTB, $R_C (t, a)$. 
We use this model to calculate the basic reproduction number, $R_o$, and project the endemic of numbers of infected and co-infected individuals.
To evaluate the consequences of co-infection for infection dynamics, we compare $R_o$ and endemic prevalence in modeled populations with one or both infections.
To explore how the individual-level consequences of co-infection influence co-infection dynamics, we explore the following individual-level processes in a sensitivity analysis: (1) the effects of prior infection with brucellosis on the rate of acquiring BTB infection, (2) the effects of prior infection with BTB on the rate of acquiring brucellosis infection, (3) the effects of co-infection on mortality rate, and (4) the effects of co-infection on birth rates.
Our model parameterization was informed by our data analysis (Table S3.1). 
As a result, it incorporates realistic age-specific transmission and mortality rates as well as data-driven estimates of the consequences of co-infection.
Furthermore, we incorporates uncertainty in the individual-level consequences of co-infection by conducting 1000 simulations, with parameter values drawn from the distributions defined in our data analysis (Table S3). \\

Model simulations and analyses were conducted in R and are publicly available \cite{gorsich_git}.

\pagebreak

\section* {Model Structure}
We modeled BTB infection as a directly transmitted, lifelong infection, with density dependent transmission \cite{jolles_interactions_2008}. 
Transmission of brucellosis was assumed to be frequency dependent because transmission occurs through ingestion of the bacteria shed in association with an aborted fetuses, reproductive tissues, or discharges during birthing (cattle: \cite{samartino_pathogenesis_1993}; bison: \cite{rhyan_pathogenesis_2009, rhyan_pathology_2001}). 
Vertical transmission of brucellosis has been experimentally demonstrated in cattle and bison (\textit{Bison bison}) \cite{plommet_brucellose_1973, fensterbank_congenital_1978}, but appears to play a relatively minimal role in transmission \cite{hobbs_state-space_2015}. 
In other species, such as Elk, vertical transmission has not been experimentally demonstrated \cite{thorne_brucellosis_1978}.
We did not consider vertical transmission because serological evidence suggests that it is also rare in African buffalo \cite{gorsich_context-dependent_2015}.
Following sero-conversion, buffalo remain infected and infectious with brucellosis for two years, following the time course of infection in cattle and bison \cite{rhyan_pathogenesis_2009, hobbs_state-space_2015, treanor_vaccination_2010}.  
Upon recovery from active infection, buffalo are assumed to be no longer infectious. 
Although persistent infections are possible, recrudescence is rare \cite{hobbs_state-space_2015}. \\

Our model incorporates age-structure to represent three features of our data analysis.  
First, juvenile buffalo suffer $3.25$-fold higher mortality rates compared to adult buffalo (SI Appendix 1, Table S1). 
Second, the rate at which buffalo acquired brucellosis was $2.42$-fold higher in early reproductive buffalo compared to adult buffalo (SI Appendix 1, Table S2). 
Third, only reproductive-aged buffalo were observed with a calf (SI Appendix 1, Fig S1).
We represent there processes by incorporating age-specific parameters. 
Table S3 defines the parameters and variables used in the model and Table S4 defines the values and ranges used.\\

We assume births, deaths, and the infection process occur continuously. 
We represent density dependent recruitment into the first age category \cite{sinclair_resource_1975} with a generalized Beverton and Holt equation \cite{getz_hypothesis_1996}. 
This form of density dependence is defined with two parameters.  
The abruptness parameter, $\phi$, defines the rate at which density dependence sets in around a characteristic density, defined by the scaling parameter, $K$. 
It results in a realistic stable age structure and relatively constant population size (Fig S3). 

The per capita birth rate has the form, \\
\begin{equation}
R(a, N(t)) = \frac{b(a)}{1 + \frac{N(t)}{K}^\phi}
\end{equation}

where $b(a)$ is the per capita age-specific birth rate at low densities and N(t) is the total population size. We did not find evidence of disease-induced reductions in fecundity (SI Appendix 1, Fig S1), such that the number of births into an age class is defined by,

\begin{equation}
r(a, N(t)) = 
\begin{cases}
\int_{0}^{\infty}  R(\alpha, N(t))N(\alpha, t) d\alpha & \text{if $a = 1.$} \\
0 & \text{otherwise} \\
\end{cases}
\end{equation}

\begin{figure}
  \centering
  \includegraphics[width = \textwidth]{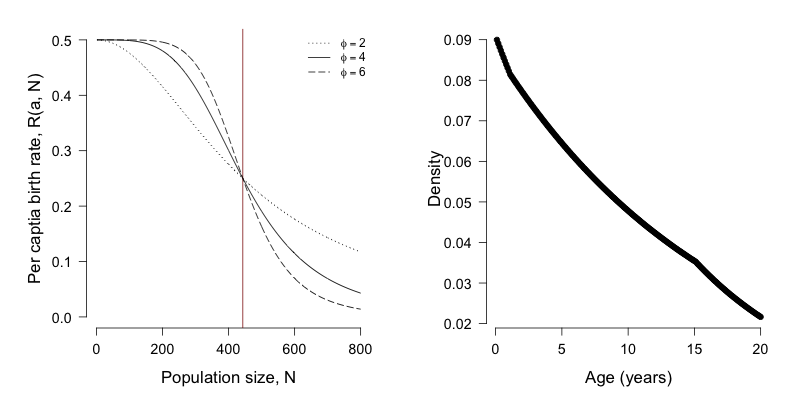}
  \caption*{\textbf{Fig S3.} (left) The per-capita birth rate decreases with increasing population size.  Increasing the abruptness parameter, $\phi$, results in stronger density dependence around, $K$ (red line). (right) The stable age distribution in the absence of infections appears visually similar to field estimates \cite{jolles_population_2007, Caron_ecological_2003}. }
\label{fig:figS3}
\end{figure}

\begin{table} 
\caption*{\textbf{Table S3.} Notation used for model variables and parameters.}
\newcommand{\head}[1]{\textnormal{\textbf{#1}}}
\small
\begin{tabular}{ll} 
\hline
\head{Symbol} & \head{Definition}\\
\hline
\textbf{Variables} &   \\
$S(t, a)$ & buffalo susceptible to both infections of age a at time t  \\
$I_T(t, a)$ & buffalo infected with BTB only of age a at time t  \\
$I_B(t, a)$ & buffalo infected with brucellosis only and infectious of age a at time t \\
$I_C(t, a)$ & buffalo co-infected with both pathogens of age a at time t  \\
$R_B(t, a)$ & buffalo persistently infected with brucellosis but no longer infectious of age a at time t  \\
$R_C(t, a)$ & buffalo persistently infected with brucellosis and co-infected with BTB of age a at time t \\
& \\
\textbf{Parameters} &   \\
$b(a)$ & age-specific maximum birth rate at low population size\\
$K $& scaling parameter defining the characteristic population size \\
$\phi $& abruptness parameter controlling the strength of density dependence around K \\ 
$\mu_S(a) $& age-specific mortality rate for susceptible buffalo \\ 
$\mu_T(a) $& age-specific mortality rate for buffalo with BTB only \\ 
$\mu_B(a) $& age-specific mortality rate for buffalo with brucellosis \\ 
$\mu_C$ &age-specific mortality rate for buffalo co-infected with both pathogens \\ 
$\beta_T $ & transmission rate for BTB in susceptible buffalo \\
$\beta_B(a)$ & transmission rate for brucellosis in susceptible buffalo \\
$\beta_{T}^{'}$  & transmission rate for BTB in buffalo with brucellosis \\
$\beta_{B}^{'}$  & transmission rate for brucellosis in buffalo with bTB \\
$\gamma$& recovery rate for brucellosis \\
$\epsilon$& recrudescence rate for brucellosis \\
\hline 
\end{tabular}
\end{table}

\begin{table} 
\caption*{ \textbf{Table S4.} Parameter values, dimensions, and references.}
\newcommand{\head}[1]{\textnormal{\textbf{#1}}}
\small
\begin{tabular}{llcc} 
\hline
\head{Parameter} & \head{Value} & \head{Dimensions} & \head{Reference}\\*
\hline
$b(a)$ & 0.5 if $a \geq 5$, 0 otherwise & $yr^{-1}$ & - \\*
$K $& 433 & indiv & \cite{cross_assessing_2006} \\*
$\phi $& 4 & dimensionless & \cite{cross_assessing_2006} \\* 
$\mu_S(a) $& 0.06 if $2 < a \leq 16$, 0.1 otherwise & $yr^{-1}$ & \cite{cross_assessing_2006, jolles_hidden_2005} \\* 
$\mu_T(a) $& $2.82 \mu_S(a)$ & $yr^{-1}$ & Table S1 \\* 
$\mu_B(a) $& $3.02 \mu_S(a)$ & $yr^{-1}$ & Table S1 \\ 
$\mu_C$ & $8.58 \mu_S(a)$ & $yr^{-1}$ & Table S1 \\ 
$\beta_T $ & $1.331 * 10 ^{-4}$ & $indiv^{-1}day^{-1}$ & fit \\
$\beta_B(a)$ & $0.576$ if $a\geq 5$, $0.576 exp(0.885)$ otherwise & $day^{-1}$ & fit, Table S2 \\
$\beta_{T}^{'}$  & $\beta_T$ & $indiv^{-1}day^{-1}$ & Table S2 \\
$\beta_{B}^{'}$  & $2.09 \beta_{B}(a)$ & $day^{-1}$ & Table S2 \\
$\gamma$& 0.5 & $day^{-1}$ & \cite{rhyan_pathogenesis_2009} \\
$\epsilon$& 0.3 & $day^{-1}$ & \cite{hobbs_state-space_2015, treanor_vaccination_2010, ebinger_simulating_2011} \\
\hline 
\end{tabular}
\end{table} 

\pagebreak

\noindent
These assumptions give the following system of partial differential equations, \\
\begin{align*}
\Big \{ \frac{\partial}{\partial t} + \frac{\partial}{\partial a} \Big \} S_{}(t, a) &= r(a, N(t)) - (\lambda_{T}(t) + \lambda_{B}(t, a) + \mu_{S}(a)) S(t, a) \\*         
\Big \{ \frac{\partial}{\partial t} + \frac{\partial}{\partial a} \Big \} I_{T}(t, a)&= \lambda_{T}(t) S(t, a) -  (\lambda'_{B}(t, a) - \mu_{T}(a)) I_{T}(t, a) \\*
\Big \{ \frac{\partial}{\partial t} + \frac{\partial}{\partial a} \Big \}  I_{B}(t, a)&=  \lambda_{B}(t, a) S(t, a) - (\lambda'_{T}(t) + \gamma + \mu_{B}(a)) I_{B}(t, a) + \epsilon R_{B}(t, a) \\*
\Big \{ \frac{\partial }{\partial t} + \frac{\partial}{\partial a} \Big \}  I_{C}(t, a)&= \lambda'_{T}(t) I_{B}(t,a) + \lambda'_{B}(t, a) I_{T}(t, a) + \epsilon R_{C}(t, a) - (\gamma + \mu_{C}(a)) I_{C}(t, a) \\*
\Big \{ \frac{\partial}{\partial t} + \frac{\partial}{\partial a} \Big \}  R_{B}(t, a)&=  \gamma I_{B}(t, a) - (\epsilon + \mu_{B}(a)) R_{B}(t, a) \\*            
\Big \{ \frac{\partial}{\partial t} + \frac{\partial}{\partial a} \Big \} R_{C}(t, a)&=  \lambda'_{T} R_{B}(t, a) + \gamma I_{C}(t, a) - (\epsilon + \mu_{C}(a)) R_{C}(t, a) \\* 
\end{align*}

\noindent
with force of infection, \\*
\begin{align*}
\lambda_{T}(t) &= \beta_T \int_{0}^{\infty} (I_t(t,\alpha) + I_{C}(t,\alpha) + R_{C}(t, \alpha)) d\alpha\\*
\lambda'_{T}(t) &= \beta'_T \int_{0}^{\infty} (I_t(t,\alpha) + I_{C}(t,\alpha) + R_{C}(t, \alpha)) d\alpha\\*
\lambda_{B}(t, a) &= \beta_{B}(a) \int_{0}^{\infty} (I_{B}(t,\alpha) + I_{C}(t,\alpha)) d\alpha\\*
\lambda'_{B}(t, a) &= \beta'_{B}(a) \int_{0}^{\infty} (I_{B}(t,\alpha) + I_{C}(t,\alpha)) d\alpha\\*
\end{align*}

We used this model to investigate the consequences of co-infection for Ro and endemic infection prevalence. 
We evaluated the consequences of co-infection by comparing infection levels in scenarios where both diseases were present to scenarios with a single infection.
We calculated endemic infection levels numerically using the method-of-lines with the ode.1D function in the deSolve package in R \cite{desolve_package}.
We calculated Ro numerically using the next generation method \cite{van_den_driessche_reproduction_2002}, reviewed in \cite{heffernan_perspectives_2005}.

\section*{Model sensitivity and uncertainty analysis }
We used Monte Carlo simulation to incorporate uncertainty in the individual-level consequences of co-infection quantified in our data analyses. 
Parameter estimates from Cox proportional hazards models are normally distributed with means and standard errors provided in tables S1 and S2. We drew 1000 random parameter values following the sampling distributions in table S5.
For each value, we quantified of Ro and endemic prevalence for both pathogens in populations with and without co-infection.  
Parameter values were drawn independently (Fig S4).
Because of this, some parameters with high brucellosis mortality and low facilitation with co-infection resulted in low values of brucellosis prevalence in scenarios with and without co-infection (Fig. 3).
This parameter space occurred 4\% of the time, resulting in a heavy density around no change in brucellosis prevalence with co-infection. 
BTB prevalence remained above 1\% for all parameters.

\begin{table}[hb]
\caption*{ \textbf{Table S5.} Parameter values and distributions for the individual-level consequences of co-infection.}
\newcommand{\head}[1]{\textnormal{\textbf{#1}}}
\small
\begin{tabular}{llcc} 
\hline
\head{Parameter} & \head{Calculation} & \head{Sampling distribution} \\*
\hline
Mortality in BTB$+$ buffalo ($\mu_T$) & $\mu_T (a) = \mu_S(a) * exp(\rho_1)$ & $\rho_1 \sim \mathcal{N} (\mu = 1.04, \sigma = 0.35)$  \\*
Mortality in brucellosis$+$ buffalo ($\mu_B$)& $\mu_B (a) = \mu_S(a) * exp(\rho_2)$ & $\rho_2 \sim \mathcal{N} (\mu = 1.11, \sigma = 0.35)$ \\*
Mortality in co-infected buffalo ($\mu_C$) & $\mu_C (a) = \mu_S(a) * exp(\rho_1 + \rho_2)$ & - \\* 
Increase in brucellosis transmission & $\beta'_B = exp(\rho_3) * \beta_B$ & $\rho_3 \sim  \mathcal{N} (\mu = 1.46, \sigma = 0.41)$\\* 
\hline 
\end{tabular}
\end{table} 

\begin{figure}[hb]
  \centering
  \includegraphics[width = 0.9 \textwidth]{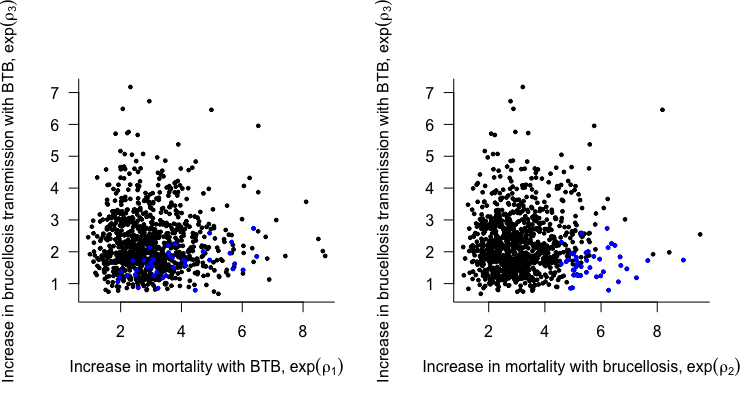}
  \caption*{\textbf{Fig S4.} Parameter values in Table S5 were drawn independently. Blue values reflect parameters resulting in a brucellosis prevalence of $0.1\%$ or less for scenarios with and without co-infection.}
\label{fig:figS4}
\end{figure}

We evaluated the sensitivity of our model to changes in the functional form of density dependence (Fig S5) and in parameter values (Fig S6, Fig S7). 
We used Latin Hypercube Sampling and partial rank correlation coefficients (PRCC) to quantify the strength of linear associations between model output and each parameter \cite{marino_methodology_2008}. Dots and error bars represent the partial rank correlation coefficients and $95\%$ confidence intervals based on 100 samples. Parameter ranges were drawn from a uniform distribution with ranges from 2 times greater than or less than the base values presented in Table S4.

\pagebreak

\begin{figure}
  \centering
  \includegraphics[width = \textwidth]{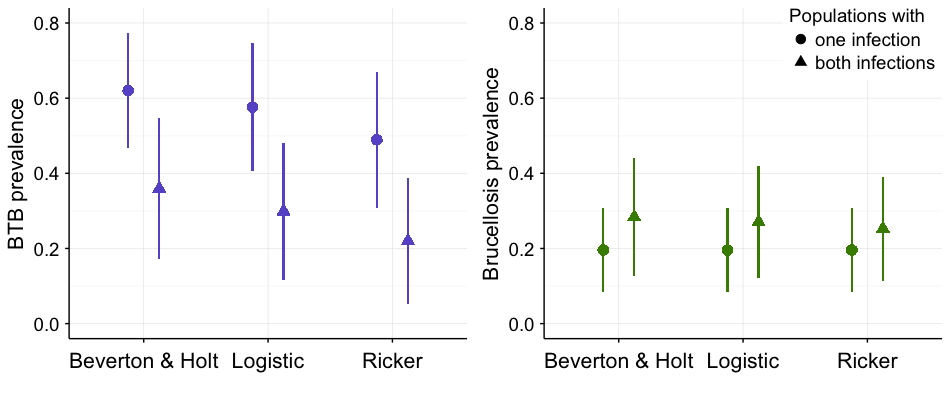}
  \caption*{\textbf{Fig S5.} Model results using alternative representations of density dependence, including logistic and Ricker representations. (a) Brucellosis prevalence in buffalo populations with and without co-infection. (b) BTB prevalence in buffalo populations with and without co-infection.  All three models result in similar qualitative patterns of infection and co-infection. Under the logistic and Ricker forms of density dependence, the per capita birth rate has the form, $R(a,N) = b(a) (1- \frac{N}{K_L} )$ and $R(a,N) = b(a)$exp$(\frac{-N}{K_R})$, respectively (15).  We set the parameter, K, such that all models produced the same population size in the absence of infection (609 individuals; $K_L = 1520.4$ for logistic; $K_R = 419.8 $ for Ricker).}
\end{figure}

\begin{figure}
  \centering
  \includegraphics[width = \textwidth]{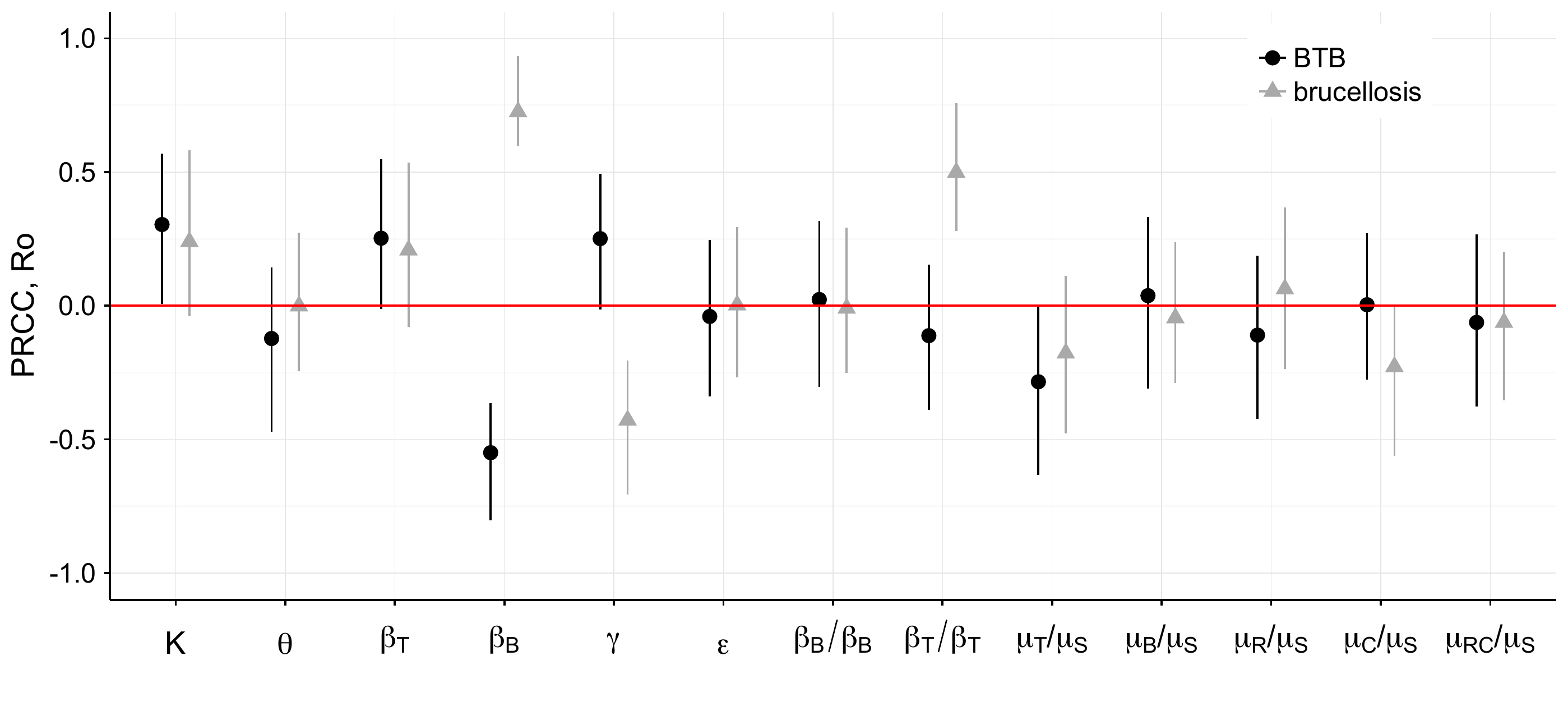}
  \caption*{\textbf{Fig S6.} Partial rank correlation coefficients and 95\% confidence intervals for $R_o$.  Colors represent the effect of a given parameter on the $R_o$ for BTB (black) or brucellosis (gray). Confidence intervals account for the 13 multiple comparisons considered here using a Bonferroni correction.}
\label{fig:figS6}
\end{figure}

\begin{figure}
  \centering
  \includegraphics[width = \textwidth]{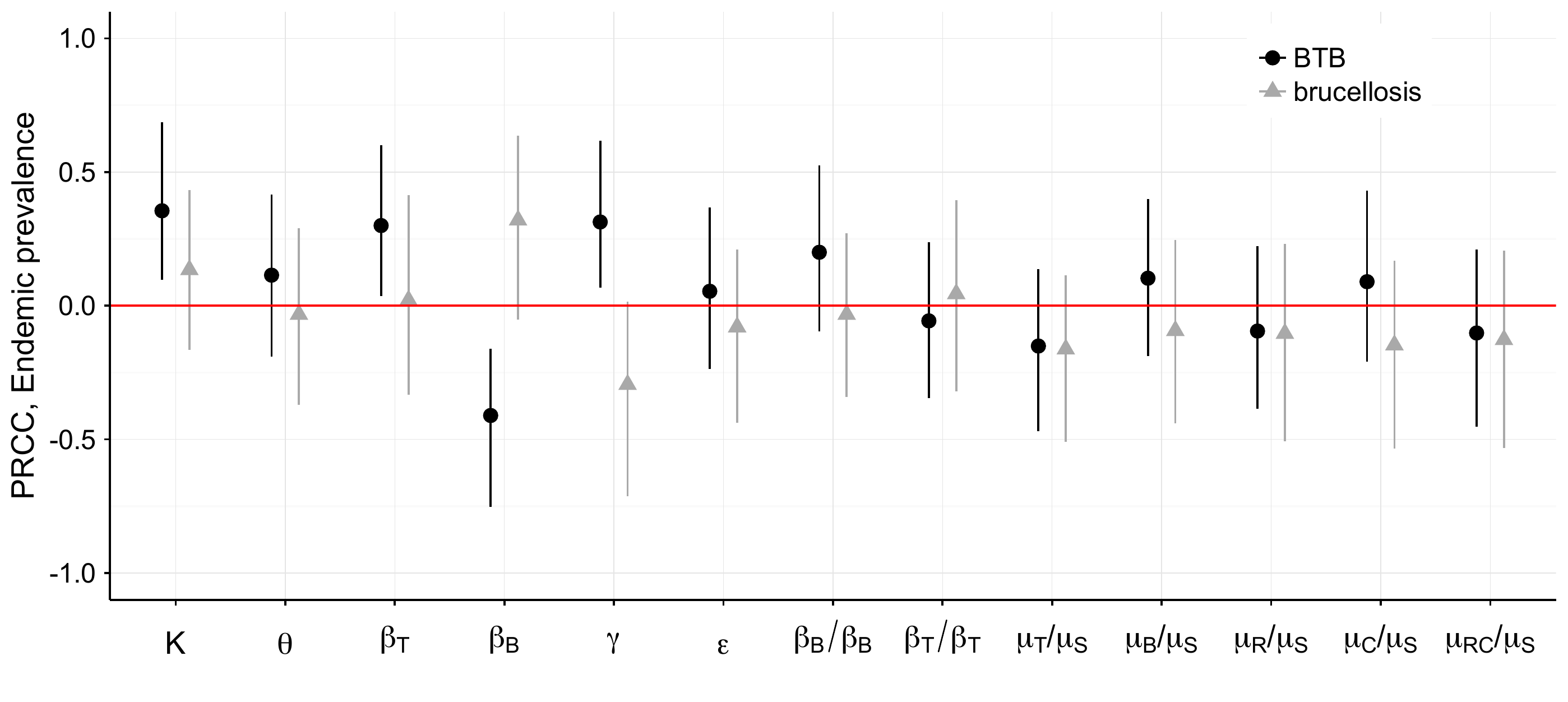}
  \caption*{\textbf{Fig S7.} Partial rank correlation coefficients and 95\% confidence intervals for endemic prevalence.  Colors represent the effect of a given parameter on the prevalence of BTB (black) or brucellosis (gray). Confidence intervals account for the 13 multiple comparisons considered here using a Bonferroni correction.}
\label{fig:figS7}
\end{figure}

\pagebreak
\clearpage

\noindent \Large{\textbf{Appendix 3. Additional information on field methods and diagnostic testing}}\\
\normalsize
Detailed capture information, including chemical immobilization methods and recapture rates have been described previously \cite{beechler_enemies_2015, gorsich_context-dependent_2015, ezenwa_opposite_2015}. We collected all field data Kruger National Park (KNP), South Africa. Data collection methods for the results presented in this work are summarized briefly here. 

\subsection*{Buffalo cohort study}
 We conducted a longitudinal study of 151 female buffalo to estimate the consequences of BTB and brucellosis infection. An initial 104 female buffalo were captured between June 2008 and October 2008. Buffalo were captured at two locations in the southeastern section of KNP. Fifty-three buffalo were captured in the Lower Sabie region; fifty-one were captured in the Crocodile Bridge region (Fig S8). Buffalo were radio-collared for re-identification and captured biannually at approximately 6 month intervals from their initial capture in 2008 until August of 2012. As natural mortalities occurred throughout the study period, new buffalo were captured and monitored, so that 47 buffalo were added throughout the course of the study.
 
During each capture, we recorded brucellosis infection status, BTB infection status, age, and the animals' reproductive status. Buffalo age was assessed by tooth eruption in younger buffalo and by incisor wear in older buffalo \cite{jolles_population_2007}. Reproductive status was quantified by whether the buffalo had a calf (defined as animals $< 1$ year old) at the capture immediately following the birthing season (March - July; \cite{gorsich_context-dependent_2015}). For each capture, we used a combination of visual siting and lactation status to identify buffalo with calves. Buffalo calves often associate with their mother until the next breeding season and were distinguishable from other calves by being within a few meters from the female. In 6\% (13/ 204) of observations, calf status was marked as unknown when a calf was observed but it was not clear who was the mother. In these cases, the focal animal was identified as having a calf only if they were lactating. \\

\subsection*{Disease testing}
Blood was collected by jugular venipuncture into lithium heparinized tubes for BTB diagnostics and into tubes with no additives for brucellosis diagnostics. Diagnosis of brucellosis was based an ELISA test (Brucellosis Serum Ab ELISA test, IDEXX). The brucellosis ELISA detects the presence of antibodies in serum and has an estimated sensitivity of 93\% and specificity of 87\% in African buffalo \cite{gorsich_evaluation_2015}. Diagnosis of BTB infection was determined based on a whole blood gamma interferon ELISA assay (BOVIGAM ELISA kit, Prionics; \cite{michel_approaches_2011}). The BOVIGAM assay for BTB identifies an infected animal based on the \textit{in vitro} production of IFN-gamma by whole blood after stimulation with \textit{Mycobacterium bovis} antigen (bovine tuberculin). Diagnosis of BTB is complicated by non-specific reactivity to environmental mycobacteria, including \textit{Mycobacterium avium} antigens. Therefore, we followed an established protocol that accounts for exposure to \textit{M. avium} because it has an estimated sensitivity of 86\% and specificity of 92\% after optimization for use in African buffalo \cite{michel_approaches_2011}. For a few captures, diagnosis was not possible if blood collection was incomplete or if errors in testing resulted in unclear test results. Removal of these capture events resulted in the removal of only six buffalo from our dataset (146 buffalo sampled).

\begin{figure}
\centering
\includegraphics[width=.99\linewidth]{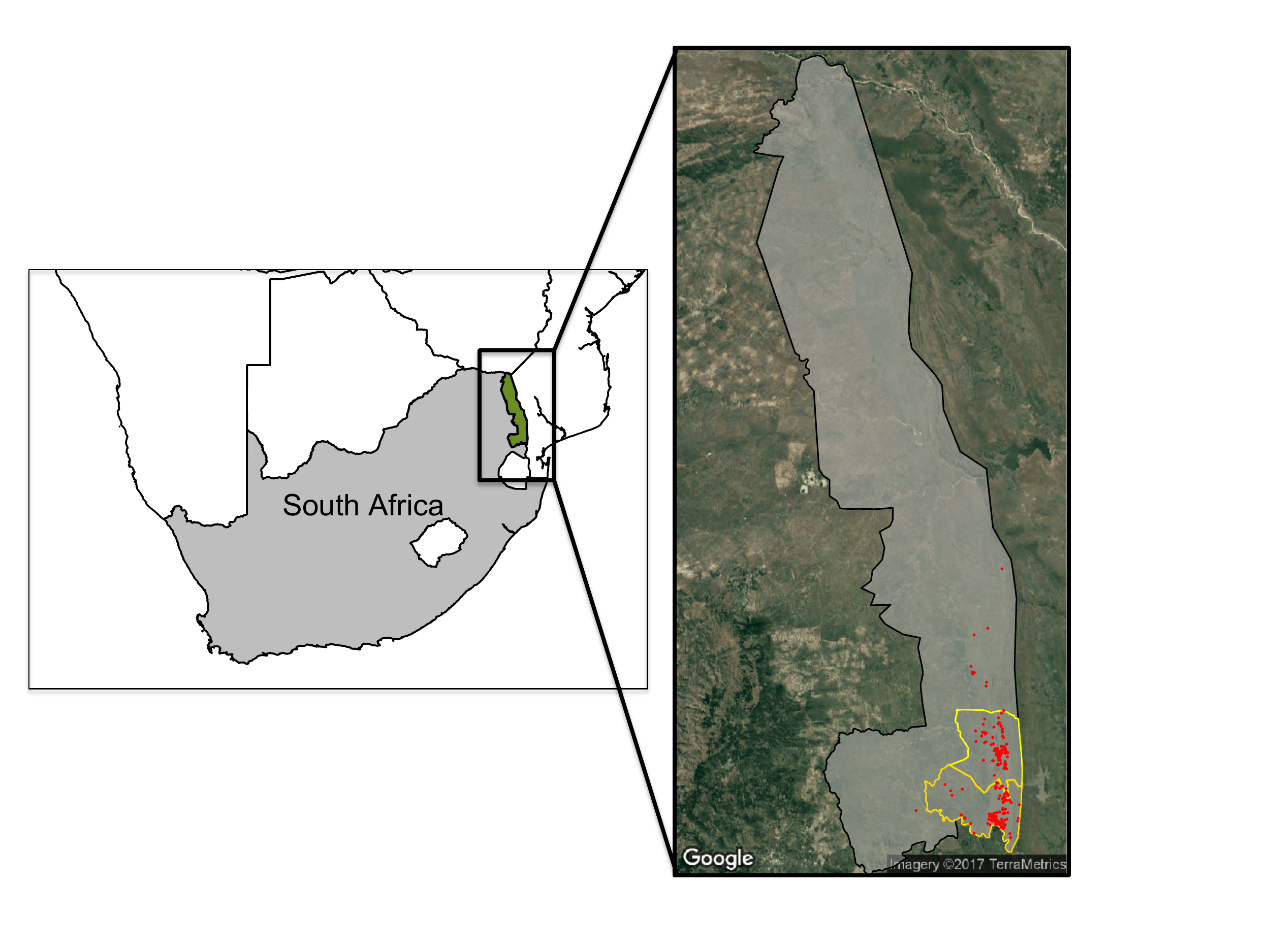}
\caption*{\textbf{Fig S8.} Kruger National Park is on the northeastern boundary of South Africa.  The blown-up figure shows a google maps image of the park. Buffalo were initially captured in the Lower Sabie (light yellow) Crocodile Bridge sections (dark yellow) of the park. Red dots represent the re-capture locations for the subsequent capture. We account for initial capture location in all statistical analyses although more detailed analyses of how buffalo move over time and space are underway (7).}
\label{fig:figS8}
\end{figure}

\clearpage
\pagebreak

\bibliography{gorsich_et_al_coinfection}

\bibliographystyle{pnas-new}

\end{document}